\documentclass{llncs}
\usepackage{graphicx}
\usepackage{amssymb,amsmath,color}
\usepackage{caption}
\usepackage{adjustbox}
\usepackage{hyperref}
\usepackage[anythingbreaks]{breakurl}
\usepackage{algorithm}
\usepackage{booktabs}
\usepackage{longtable}
\usepackage{tabu}
\usepackage{algorithmicx}
\usepackage{soul}
\usepackage{xspace}

\newcommand{\comment}[1]{}

\renewcommand{\paragraph}[1]{\vspace{5pt}\noindent\textbf{#1\quad}}

\begin{document}
\pagenumbering{gobble}
\pagestyle{plain}
\thispagestyle{empty}

\title{Physical Cryptographic Signatures for Absentee Ballots}

\newif\ifblind\blindfalse 

\author{Matthew Bernhard\inst{1}}
\institute{Department of Computer Science and Engineering\\
      University of Michigan\\2260 Hayward
      Street, Ann Arbor, MI 48109-2121\\\url{matber@umich.edu}}
\maketitle

\begin{abstract}
    Physical signature verification on absentee ballots became a major
    flashpoint in the 2018 midterm elections in the United States, especially
    in states like Georgia, Florida, and Arizona, where close election margins
    resulted in heightened attention to the counting of absentee ballots. As
    vote-by-mail solutions are becoming more prevalent across the U.S., these
    issues are sure to continue affecting elections in the United States.
    Signature verification is an inexact science; often times guidelines can
    vary widely from jurisdiction to jurisdiction. In this paper we provide a
    cryptographic remedy to this solution that is usable, secure, and easily
    integrated into existing election infrastructure. 
\end{abstract}

\section{Introduction}

The 2018 midterm elections in the United States saw numerous contentious races
come down to the wire as jurisdictions counted and recounted ballots. One of
the biggest pressure points was that of absentee ballots: in many states,
there was intense litigation over which absentee ballots could be counted in
the totals. Georgia saw a federal court overrule the policies in Gwinnett
County, where ballots from Asian-American voters were being rejected at
significantly higher rates than other voters~\cite{gwinett2018absentee}. Other
states like Florida and Arizona saw similar
issues~\cite{florida2018signatures,arizona2018absentee}

\begin{figure}
    \centering
    \includegraphics[width=.5\columnwidth]{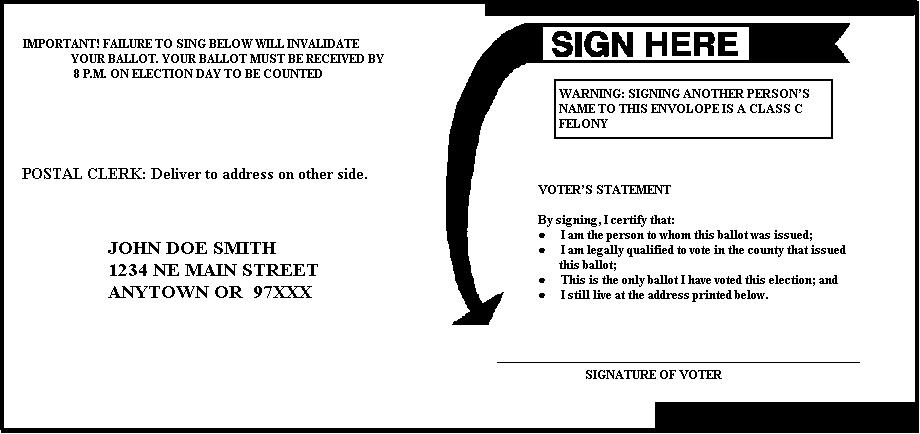}
    \caption{\textbf{Absentee Ballot Envelope from Washington County, 
    Oregon}---An example of the envelope a voter signs when mailing in 
    their ballot~\cite{washington2018vbm}.}
    \label{fig:envelope}
\end{figure}

Much of the litigation around absentee ballots focused on signatures: in the
United States, absentee ballots require the voter to sign their name to the
ballot to provide authentication and legally swear that they are not committing
fraud. An example envelope is shown in Figure~\ref{fig:envelope}. This issue is
sure to be compounded by the rise of vote-by-mail in the U.S.: in 2018 three
U.S.\ states cast ballots almost entirely absentee, and numerous other states
permit no-reason absentee voting, with the number increasing each election
cycle as more states pass election reforms~\cite{verifiedvoting}.

Signature validation is often done by hand~\cite{npr2018signatures}, though
there are automated solutions available~\cite{runbeck}. Both schemes have
inherent flaws, as human-generated signatures change over time and often are
not consistently reproduced by the signee.\footnote{The field of forensics has
numerous studies about the failure of trained experts and automated systems to
correctly identify correct or forged signatures, for
example~\cite{sita2002forensic,found1999development,bird2010forensic,harralson2014developments}.}
They can also be omitted entirely, by mistake.  Some difficulties are well
documented in election worker training manuals, such as~\cite{cosigs} from
Colorado. 
Local regulations and election worker training may change between
jurisdictions, so absentee ballot rejection rates based on signature validation
can vary by several percentage points even within one state~\cite{aclu2018fl}.
All of these factors contribute to a need for a more robust method of voter
authentication for absentee ballots.

In this paper, we present an application of cryptography to solve this problem.
Local governments can establish a shared secret with voters through processes
which are already in place (voter registration), with little additional cost.
Voters can then use this shared secret to derive a cryptographic signature code
for their ballot, which they will write on the envelope in place of a
handwritten signature bearing the voter's name. 

Using signature codes like 14- or 20-digit numbers is a vast improvement over
handwritten name-signatures, as there is less room for interpretation of a
voter's handwriting. Techniques like OCR can be more effectively applied to
this approach, as the form of the signature is no longer important, only its
content. 

Asking voters to write down long numbers may not be an acceptably usable
approach. We propose some mitigations to this process, and argue that writing
signatures codes in this way is not inherently less usable than the current
standard. Handwritten name-signatures can also always be used as a backup, in
case the technology fails or a voter has difficulty using the approach. 

The rest of this paper is structured as follows: Section~\ref{sec:background}
provides an overview of the current vote-by-mail landscape and the role
signatures play, as well as providing a threat model.
Section~\ref{sec:implementation} develops our approach and addresses some of
the nuance therein. Finally, Section~\ref{sec:discussion} contextualizes the
solution and concludes. 

\section{Background}
\label{sec:background}

\subsection{How absentee voting works}

Absentee ballots in the United States function in one of two ways. Some states,
like Colorado, Washington, and Oregon, send ballots to all registered voters a
few weeks prior to the election. Other states, like Michigan, Texas, and
Georgia, require voters to explicitly request absentee ballots. Ballots are
sent via the postal service, and voters have up until some deadline to fill
them out and mail them back to their election office. 

A typical absentee ballot voting process involves filling out the ballot,
placing the ballot in an inner secrecy envelope, and then putting the secrecy
envelope with the ballot inside an outer envelope. The voter then signs the 
outer envelope and mails the ballot back to their clerk.

\subsection{Threat Model}

Postal voting in this way has four main potential adversaries:

\begin{enumerate}
    \item {\bf Coercers}---coercers are often people the voter knows, e.g.\ a
        spouse or boss. They can cause (either by force or by enticement) the
        voter to surrender her ballot and either vote in the voter's place, or
        simply disenfranchise the voter.  Coercers may forge the voter's
        signature or compel the voter to sign. Signature validation may detect
        this kind of attack, if the coercers do not produce faithful
        forgeries~\cite{sentinel2016,charlotte2018fraud}.
    \item {\bf Impersonators}--these voters order ballots on behalf of
        people who don't vote frequently (or who are no longer eligible to vote
        but have not been removed from the voter rolls. There is little to no
        evidence that this type of election fraud occurs in the
        U.S.~\cite{brennan2017debunking}.
    \item {\bf Insiders}---insiders run the elections. They can potentially
        reject ballots or change the votes on a ballot, assuming there are not
        strong policies in force to prevent this. This kind of issue has been
        raised by litigation about ``exact match,'' where election officials
        use very strict rules when evaluating voter information and
        signatures~\cite{augusta2018exact}.
    \item {\bf Nation-states}---nation-states can prevent ballots from being
        delivered to the voter by sending a flurry of fake absentee requests
        for the wrong address, or by meddling in voter registration data. We
        have already seen intrusion into voter registration systems in the
        United States~\cite{ssci18}. They may also be able to tamper with
        tabulation of the ballots, but this is outside our scope. 
\end{enumerate}

It is worth pointing out that the current signature validation scheme used in
much of the U.S.\@ provides no mitigation to any of these threats. There are
other procedural security measures that can prevent, say, insiders from
throwing out ballots or nation-states from overwhelming the postal system, but
we shall consider them outside of scope. 

\section{Implementation}
\label{sec:implementation}

\begin{figure*}
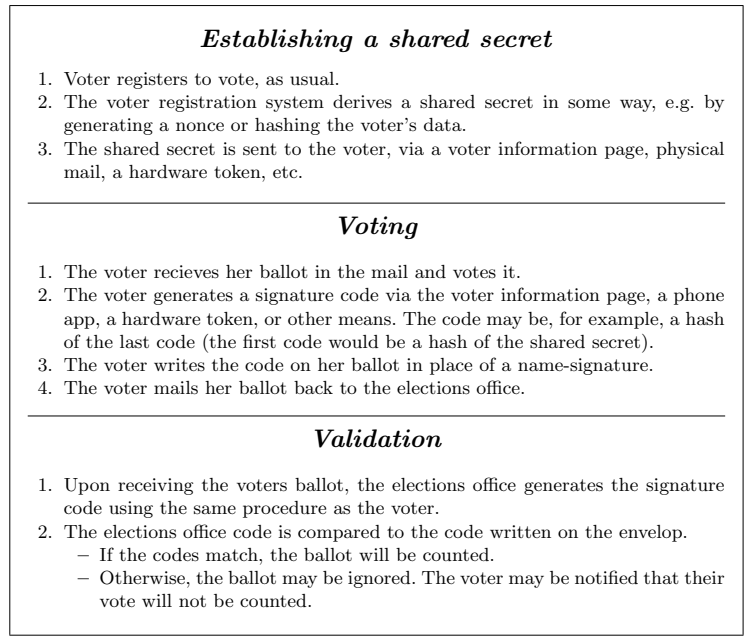

    \centering
\scalebox{.8}{
\framebox[\textwidth]{
  \parbox{0.95\textwidth}{
    \vspace{8pt}
    \centerline{\bf\emph{\large Establishing a shared secret}}
    \vspace{-3pt}
\begin{enumerate}
    \item Voter registers to vote, as usual.
    \item The voter registration system derives a shared secret in some way,
        e.g.\ by generating a nonce or hashing the voter's data. 
    \item The shared secret is sent to the voter, via a voter information page,
        physical mail, a hardware token, etc.
\end{enumerate}
\hrule\medskip
\centerline{\bf\emph{\large Voting}}
\begin{enumerate}
    \item The voter recieves her ballot in the mail and votes it. 
    \item The voter generates a signature code via the voter information page,
        a phone app, a hardware token, or other means. The code may be, for
        example, a hash of the last code (the first code would be a hash of the
        shared secret). 
    \item The voter writes the code on her ballot in place of a name-signature. 
    \item The voter mails her ballot back to the elections office. 
\end{enumerate}
\hrule\medskip
\centerline{\bf\emph{\large Validation}}
\begin{enumerate}
    \item Upon receiving the voters ballot, the elections office generates the 
        signature code using the same procedure as the voter. 
    \item The elections office code is compared to the code written on the 
        envelop. 
    \begin{itemize}
        \item If the codes match, the ballot will be counted. 
        \item Otherwise, the ballot may be ignored. The voter may be notified
            that their vote will not be counted. 
    \end{itemize}
\end{enumerate}
}
}
}
\normalsize
\caption{\bf Step-by-step procedure for implementing signature codes}
\label{fig:procedure}
\end{figure*}

In this section we discuss our scheme for absentee voter authentication and
some of the implementation details. A step-by-step procedure is provided in 
Figure~\ref{fig:procedure}.

\subsection{The Scheme}

The first critical components to our scheme is a shared secret established
between the local or state elections office and the voter. This secret can be
established at the time of registration or after the fact, and may be derived
from information about the voter's data as well as a source of randomness.
Details about this shared secret and its derivation are further discussed in
Section~\ref{subsec:secret}.

The other component in our scheme is a signature code generator, which takes
the voter's shared secret as input and produces a unique identifier that can
only be generated by the voter and the elections office. Details of this
generator are further discussed in Section~\ref{subsec:gen}.

Once the voter has voted their ballot and sealed it within the envelope, they
use the generator and their shared secret to generate a value, and write it on
the envelope. Once the election office receives the envelope, they use the same
information to derive the signature code and count the ballot if it matches.
Otherwise, they reject.\footnote{We omit solutions for ``curing'' invalid
signature codes, as they are frequently highly constrained by local regulation
and outside our scope.}

\subsection{The Shared Secret}
\label{subsec:secret}
Driver's license numbers are frequently used as authentication tokens by voter
registration systems (e.g.\ in Washington, Michigan, Georgia, etc.\@). However,
driver's license numbers in some states are derived in a deterministic fashion.
This means that anyone with basic information about the voter (name, date of
birth, home address, etc.\@) can derive the voter's driver's license number and
potentially change the voter's registration data. Utilities to do this sort of
thing already exist on the web~\cite{highprogrammer}. 

Another authentication token used by existing voter registration systems is the
last four digits of a social security number, a national identifier used for
administrative purposes and authentication by numerous entities within the
United States.  These numbers are also deterministic, and do not alone make a
good token for authentication. Worse, these data regularly leak due to data
breaches~\cite{forbesvoter}, and moreover most voter data is already publicly
available~\cite{igielnik2018commercial}.

Jurisdictions should thus use some other form of data to derive the shared
secret with the voter. One option would be a hash of the voter's existing voter
information data and a random nonce securely generated by the voter
registration system. Voters can then regenerate the secret as needed, and the
secret can be updated on a per-election basis and when the voter updates his or
her information. If the voter discloses her secret in any manner, the situation
can be remedied by generating another secret. 

It is likely desirable for the secret generation process to be controlled by
the election office, as an attacker with the voter's registration information
may be able to generate their own shared secret and impersonate the voter.
This attack is already present in the current signature validation model as an
attacker could forge the voter's signature or coerce the voter into signing a
ballot they control, so we consider this attack out of scope.  Another
consideration is that the transport mechanism of the shared secret, once
generated, must be protected, otherwise an attacker could simply sniff the
secret and impersonate the voter. 

\subsection{Generating a signature code}
\label{subsec:gen}

Signatures can be generated in myriad ways. A pseudo-random number generator
can be used, with the shared secret as a seed. Alternatively, a hash of the
secret can be used. 

Of course, if the shared secret is established as a traditional cryptographic
key, say, an RSA key, then the signature code can also be a cryptographic
signature signed over some pre-established information (e.g.\ the voter's name,
the date of the election, or in electronic voting schemes, the content of the
ballot).  

The generation of signature codes can be done online or offline, and we
encourage the process to be done in a transparent manner. For most voters, it
is probably sufficient to embed signature codes onto voter information
websites, or distribute codes via other mechanisms like a phone app or text
message. Some voters, like those overseas or in the military, may not have
sufficient access to derive codes in this way. To handle this, alternative
applications may be feasible, like hardware tokens with the shared secret and
generator function embedded in them. In the United States, the military has
some infrastructure to support this already, like the Common Access
Card~\cite{cac}.

\subsection{A realistic implementation}

\begin{figure}
    \centering
    \includegraphics[width=.5\columnwidth]{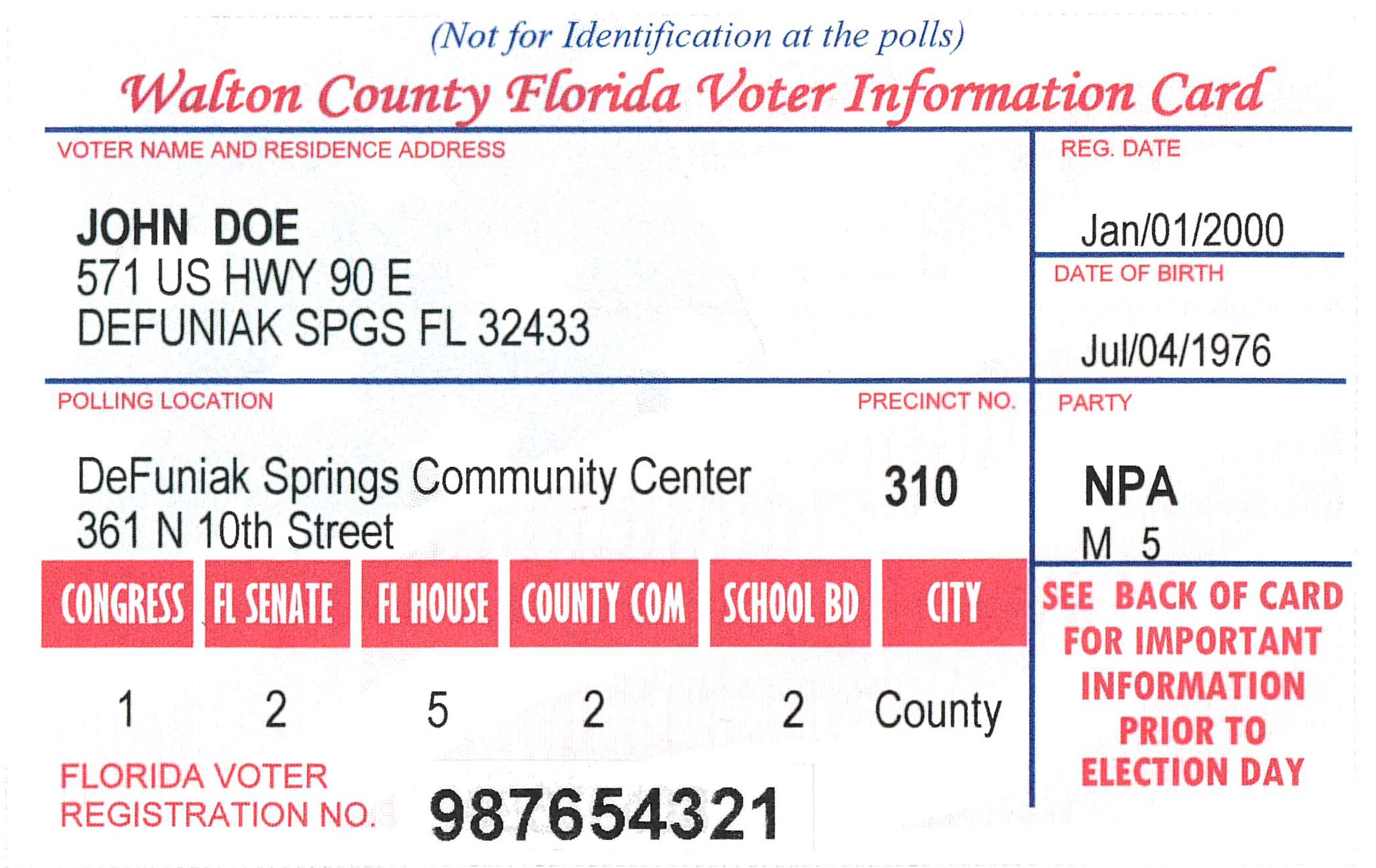}
    \caption{\textbf{Voter information card from Walton County, Florida}---a
    sample voter information card, which is sent to voters with their precinct
    and ballot style information after registering to vote and before each
    election. This card can serve as the mechanism for distributing the shared
    secret (and in fact the voter ID might be usable as the secret). Taken
    from~\cite{flvoterinfo}.}
    \label{fig:votercard}
\end{figure}

A likely implementation of our scheme is that the voter and election office
agree upon a shared secret during registration, and that the voter registration
system provides the functionality to generate signatures code for the voter.
This can be done with just a hash function: provided the voter authenticates
herself to the voter registration system, each signature code can be just a
hash of the shared secret and a nonce or timestamp, or a hash chain of past
signature codes. 

This implementation solves the problem of faulty signature code validation, and it
does not subject voters to more risk from bad actors than the current
handwritten name signature system. Further, if the system is implemented such
that the voter can increment the code at will, this may provide some weak
protection against a coercer. E.g., if a voter has her ballot submitted by a
coercer, she can increment the signature code such that when the elections
office receives her ballot, it will be canceled due to an expired signature code.
This is not unlike the alarm code approach hinted at
in~\cite{bernhard2017public}, and approaches like Civitas~\cite{civitas} and
Juels, Catalano and Jakobsson~\cite{juels05:e-vote}.

This approach does not prevent against nation-state attacks or insider threats,
however it can mitigate coercion and impersonation, which is an improvement
over the current signature-based system. 

\begin{figure}
    \centering
    \includegraphics[width=.7\columnwidth]{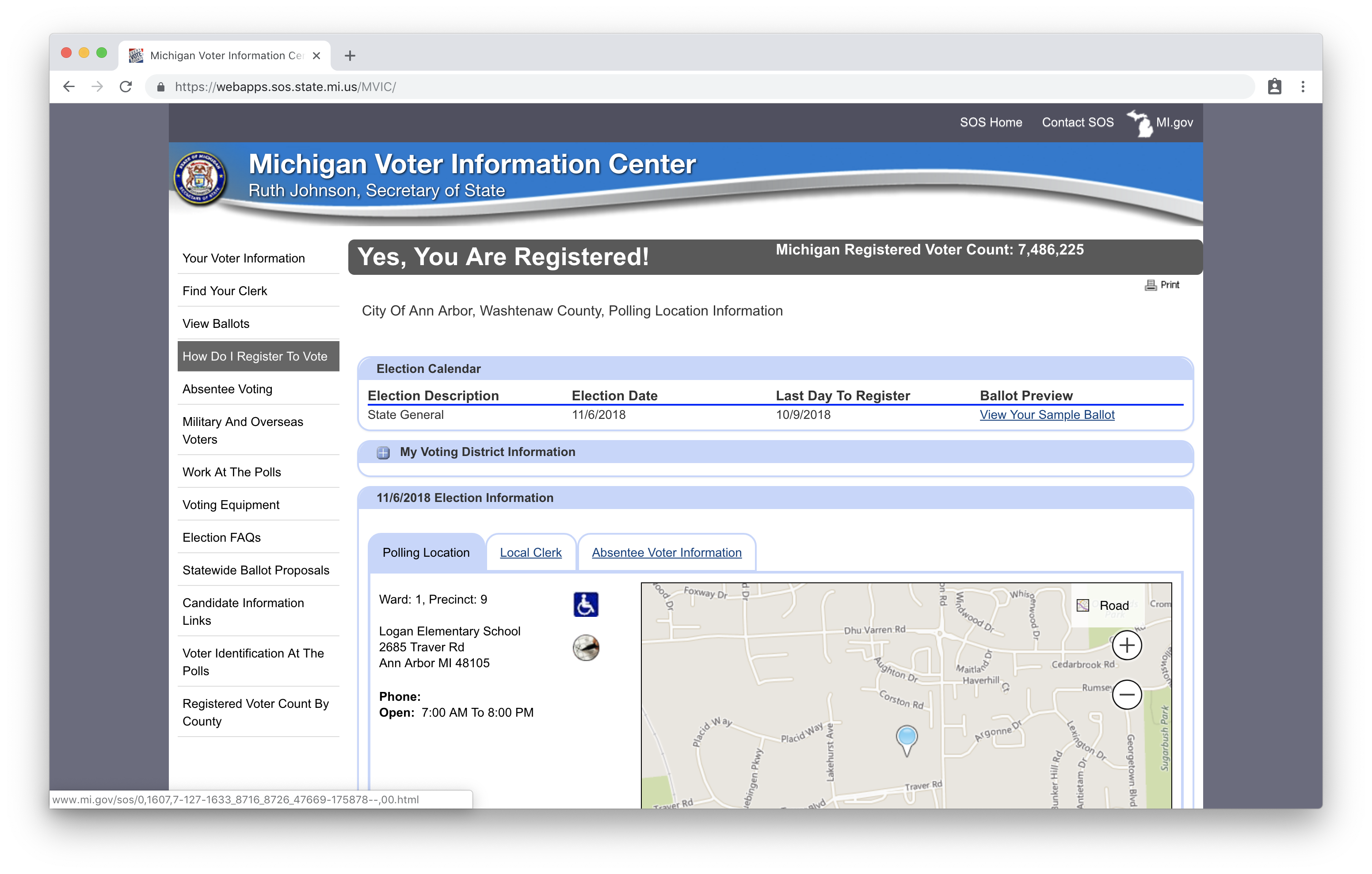}
    \caption{\textbf{Michigan Voter Information Page}---signature codes
    can be generated via a voter page like this one.}
    \label{fig:mvp}
\end{figure}

This implementation is conceptually simple and easy to implement from an
engineering perspective. Most U.S.\ states already have online voter
information portals (for instance, see Figure~\ref{fig:mvp}), and adding a hash
output to the voter information page would be a fairly simple task. Further,
states already communicate out-of-band with voters, sending voter registration
cards through the mail (an example can be seen in Figure~\ref{fig:votercard}.
The shared secret could be generated and printed on the card. Ultimately,
this solution is cheap and simple.

\section{Discussion}
\label{sec:discussion}

In this section we discuss some properties of our scheme, suggest where it can
be applied, and discuss some future work. 

As a brief aside, this system prevents malicious or intentional
disenfranchisement of voters due to strict pattern matching during the voter
registration process. If strict matching is used, this scheme can provide
additional assurance, or act as an alarm of sorts,  that a mismatch was
unintentional and warrants further investigation by an election official 

\subsection{Related Work}

Internet voting systems like the in Estonia~\cite{estonia,estoniareport} and
the experiments in Norway~\cite{norway,norway-carter,Gjosteen2016} may benefit
from this scheme. Notably, Estonia already provisions a key pair for its
citizens in its national ID, and this key pair might provide stronger
properties to our scheme. 

Other end-to-end voting systems, like those described in Bernhard et
al.~\cite{bernhard2017public} may also benefit from a more robust voter
authentication protocol. Remotegrity~\cite{DBLP:conf/acns/ZagorskiCCCEV13} was
an adaptation of the Scantegrity II voting system to a mail ballot and online 
voting system.

\subsection{Usability}

Once generated, the voter still has to correctly transcribe the signature  code
onto the ballot envelope. This may present a usability issue: human
transcription of many-digit numbers is not well
studied.\footnote{\cite{little2009turkit} provides some discussion of
handwriting recognition, that's somewhat unrelated. There has also been some
work on CAPTCHAs, which is a similar task~\cite{bursztein2010good}}. If this
does present an issue, other kinds of signature codes may be generated.
Following after some of the usable password work~\cite{bonneau2012quest},
signatures codes may be random sentences constructed based on the output of the
signature code function. This provides a neat property: even if the voter makes
a mistake, it is unlikely that the mistake will make their signature code
unverifiable: a typo in one word of a sentence does not obscure the content of
teh sentence. 


Further work is needed to establish what impact our scheme has on the usability
of absentee voting. Better understanding how accurate humans are at
transcribing numbers is paramount to making our scheme work. Moreover,
requiring voters to perform an additional step in the process voting may have
negative affects on voters' ability to vote successfully, and this also needs
further study.

\subsection{Conclusion}

In this paper we presented a remedy to the problem of absentee ballot signature
validation. Signature validation will only become a more pressing problem as
absentee voting expands in the United States and elsewhere. This paper is a
first step towards solving voter authentication for remote voting, and we hope
that future work will illuminate better mechanisms to solve this problem.

{\footnotesize
\bibliography{paper}
\bibliographystyle{abbrv}
}

\end{document}